\documentclass[11pt,a4paper]{article}
\usepackage{amsmath,amssymb}
\usepackage{epsfig,graphicx}

\def\starup#1{\mbox{$\raise1.6ex\hbox{$*$} \kern-0.5em#1$}}
\def\starupp#1{\mbox{$\raise1.8ex\hbox{$*$} \kern-1.0em#1$}}
\def\staruppp#1{\mbox{$\raise1.0ex\hbox{$*$} \kern-0.5em#1$}}
\def\sstarup#1{\mbox{\scriptsize $\raise1.8ex\hbox{$*$} \kern-.7em#1$}}

\def\lineup#1{\mbox{$\raise1.0ex\hbox{--} \kern-0.6em#1$}}

\begin{document}

\title{\bf On mass limits for leptoquarks from $ K_L^0 \to e^{\mp} \mu^{\pm} , \; 
 B^0 \to e^{\mp} \tau^{\pm} $ decays }
\author{ A.~D.~Smirnov $^a$\footnote{{\bf e-mail}: asmirnov@univ.uniyar.ac.ru}
%C.~D.~Author$^{a}$\footnote{{\bf e-mail}: cdauthor@institution.domain}
\\
$^a$ \small{\em Yaroslavl State University } \\
\small{\em Yaroslavl, Russia}
}
\date{}
\maketitle

\begin{abstract}
The contributions of the scalar leptoquark doublets into widths of
the $K^0_L \to e^{\mp} \mu^{\pm}$, $B^0 \to e^{\mp} \tau^{\pm}$
decays are calculated in MQLS model with Higgs mechanism of the
quark-lepton mass splitting. The resulted  mass
limits for the scalar leptoquarks are investigated in comparision 
with those for vector and chiral gauge leptoquarks. It is shown 
that the scalar leptoquark mass limits are essentially weaker (about $200 \, GeV$) 
than those (a few hundreds $TeV$) for gauge leptoquarks. 
The search for such scalar leptoquark doublets at LHC and the search for 
the leptonic decays $ B^0 \to l^+_i l^-_j $ are of interest.     
\end{abstract}

The search for a new physics beyond the Standard Model (SM) is now one
of the aims of the high energy physics. One of the possible
variants of such new physics can be the variant induced by the
possible four color symmetry \cite{PS} between quarks and leptons.
The four color symmetry can be unified with the SM by the gauge group 
\begin{eqnarray}
     G_{new}=G_c \times SU_L(2) \times U_R(1) 
\label{eq:Gnew}
\end{eqnarray}
where as the four color group $G_c$ can be the vectorlike group \cite{PS, AD1, AD2}   
\begin{eqnarray}
    G_c = SU_V(4) 
\label{eq:G4V}
\end{eqnarray}
the general group of the  chiral four color symmetry  
\begin{eqnarray}
G_c = SU_L(4) \times SU_R(4)  
\label{eq:G4LR}
\end{eqnarray}
or  one of the special groups  
\begin{eqnarray}
      G_c = SU_L(4) \times SU_R(3) , \,\,\,\, G_c = SU_L(3) \times SU_R(4)  
\label{eq:G4SLR}
\end{eqnarray}
of the left or right four color symmetry. 
According to these groups the four color symmetry predicts the vector gauge leptoquarks 
 the left and right gauge leptoquarks and the left (or right) gauge leptoquarks respectively. 
The most stringent mass limits the vector leptoquarks are resulted from
$K^0_L \to e^{\mp} \mu^{\pm}$ decay and are known to be of order of $10^{3} \,\, TeV$ 
\cite{VW, KM1,KM2}. 
The four color symmetry allows also the existence of scalar leptoquarks and such particles 
have been phenomenologically introduced in ref.~\cite{BRW} 
and have been discussed in a number of papers. 
The experimental lower mass limits for the scalar leptoquarks from their direct search 
 depend on additional assumptions and are about $ 250 \, GeV $ or slightly less~\cite{PDG04}.   

It should be noted that in the case of Higgs mechanism of the quark-lepton mass splitting 
 the four color symmetry predicts \cite{AD1, AD2, ADPv} the scalar leptoquarks 
\begin{eqnarray}
S^{(\pm)}_{a \alpha} = \left ( \begin{array}{c}
S_{1 \alpha}^{(\pm)}\\
S_{2 \alpha}^{(\pm)}
\end{array} \right ).   
\label{eq:SpSm}
\end{eqnarray}
with doublet structure under electroweak group $SU(2)_L$.   
Due to their Higgs origin these scalar leptoquark doublets interact with the fermions with 
coupling constants which are proportional to 
the ratios $m_{f}/ \eta $ of the fermion masses $m_f$ to the SM
VEV $\eta$. As a result these coupling constants are small for the ordinary $u-, d-, s-$ quarks 
(~$ m_u/ \eta \sim m_d/ \eta \sim 10^{-5},  m_s/ \eta \sim 10^{-3}$~) 
but they are more significant for $c-, b-$ quarks and, especially,
for $t$-quark 
( $ m_c/ \eta \sim m_b/ \eta \sim 10^{-2} , \, m_t/ \eta \sim 0.7$ ). 
 
The analysis of the contributions of these scalar leptoquark doublets into 
rediative corrections $S-, T-, U-$ parameters showed \cite{AD3,PovSm2} 
that these scalar leptoquarks can be relatively light, with masses below 1 TeV.   
It is interesting now to know what are the mass limits for these scalar leptoquarks 
which are resulted from $K^0_L \to e^{\mp} \mu^{\pm}$ decay and from other decays of such type. 

In this talk I discuss the mass limits for the  scalar leptoquark doublets 
which are resulted from 
$K_L^0 \to e^{\mp} \mu^{\pm} ,\; B^0 \to e^{\mp} \tau^{\pm} $ decays  
in comparision with the corresponding  mass limits for the vector leptoquarks 
in frame of MQLS - model and with those for the chiral gauge leptoquarks. 

The interaction of the gauge and scalar leptoquarks with 
down fermions can be described in the model independent way 
by the lagrangians 
\begin{eqnarray}
  \emph{L}_{Vdl} &=& (\bar{d}_{p \alpha} [ (g^L_k)_{pi}\gamma^{\mu}P_L +  
(g^R_k)_{pi}\gamma^{\mu}P_R ] l_i) V^k_{\alpha \mu} + h.c. ,
\label{eq:lagrVdl}\\
  \emph{L}_{Sdl} &=& (\bar{d}_{p \alpha} [ (h^L_m)_{pi} P_L +  
(h^R_m)_{pi} P_R ] l_i) S_{m {\alpha}} + h.c. ,
\label{eq:lagrSdl}
\end{eqnarray}
where 
$(g^{L,R}_k)_{pi}$ and $(h^{L,R}_m)_{pi}$ are the phenomenological coupling constants, 
$p, i = 1,2,3, $... are the quark and lepton generation idexes, 
the indexes $ k, m $ numerate the gauge and scalar leptoquarks, 
$\alpha=1,2,3$ is the $SU(3)$ color index and  
$P_{L,R}=(1\pm \gamma_5)/2$ are the left and right operators of fermions. 

The MQLS - model ( with $ V^1_{\alpha \mu} \equiv V_{\alpha \mu}$, 
$ V_{\alpha \mu}$ is the vector leptoquark of the model ) gives 
for the  phenomenological coupling constants 
  in~(\ref{eq:lagrVdl}),~(\ref{eq:lagrSdl}) 
the expressions 
\begin{eqnarray}
&&(g^{L,R}_1)_{pi}= \frac{g_4}{\sqrt{2}}(K^{L,R}_2)_{pi}, 
%\nonumber \\
\label{eq:gconst}
\\
&&(h^{L,R}_m)_{pi}=  (h^{L,R}_{2m})_{pi}= h^{L,R}_{pi} c^{(\mp)}_m , 
\label{eq:hmconst}
\end{eqnarray}
where $g_4 = g_{st}(M_c)$ is the $SU_V(4)$ gauge coupling constant related to the strong coupling constant   
at the mass scale $M_c$ of the $SU_V(4)$ symmetry breaking and   
\begin{eqnarray}
h^{L,R}_{pi} = -\sqrt{ 3/2} \frac{1}{\eta \sin\beta}
\Big [  m_{d_p} (K_2^{L,R})_{pi} -
(K^{R,L}_2)_{pi}m_{l_i} \Big ] ,
\label{eq:hconst}
\end{eqnarray}
$\eta$ is the SM VEV, $\beta$ is the two Higgs doublet mixing angle of the model, 
$m_{d_p}, m_{l_i}$ are the quark and lepton masses 
and  
$c^{(\mp)}_m$ are the scalar leptoquark mixing parameters enterring in superpositions 
\begin{eqnarray}
S_{2}^{(+)}&=&\sum_{m=0}^3 c_m^{(+)}S_m, \; \; \; \; \; \;
 \starup{S_2^{(-)}}=\sum_{m=0}^3 c_m^{(-)}S_m 
\label{eq:mixS}
\end{eqnarray}
of three physical scalar leptoquarks $S_1$, $S_2$, $S_3$ with
electric charge 2/3 and a small admixture of the Goldstone mode
$S_0$ (in general case
the scalar leptoquarks $S_{2 \alpha}^{(+)}$ and
$\starup{S_{2\alpha}^{(-)}}$
with electric charge 2/3 are mixed and can be
written as superpositions~(\ref{eq:mixS}) ). 
The matricies $K_a^{L,R}, a=1,2$ describe the (down for $a=2$) fermion mixing 
in the leptoquark currents and in general case they can be nondiagonal. 

In particular case of the two leptoquark mixing the superpositions~(\ref{eq:mixS}) 
can be approximately written as  
\begin{eqnarray}
S^{(+)}_2 = c S_1 + s S_2, 
\nonumber \\
\starup{S^{(-)}_2} = -s S_1 + c S_2  
\label{eq:mixS2}
\end{eqnarray}
where $c = cos\theta, s = sin\theta, $ $ \theta $ is the scalar leptoquark mixing angle.

Omitting the details of calculations we present the final expressions for the widths  
of the $K_L^0 \to e \mu , \; B^0 \to e \tau $ decays with account of the contributions 
of the vector leptoquark $ V_{\alpha \mu}$  and of the scalar leptoquarks $ S_1, S_2 $  
 with mixing~(\ref{eq:mixS2}) in MQLS - model for 
$K_2^{L}=K_2^{R}=I$ 
\begin{eqnarray}
&&\Gamma(K^0_L \to e \mu ) = 
\frac{m_{K^0} f_{K^0}^2}{64\pi} \bigg (1-\frac{m_{\mu}^2}{m_{K^0_L}^2} \bigg )^2 
\Bigg \{ \bigg \lbrack  \frac{4 \pi\alpha_{st}}{m_V^2} ( - \, \lineup{m}_{K^0}+m_{\mu}/2) - 
\nonumber\\
&& - \frac{h_1 h_2}{2} \bigg ( m_{\mu} \langle \frac{1}{m^2_S} \rangle^{L} -
 \,  \lineup{m}_{K^0} c s (\frac{1}{m_{S_1}^2} -  \frac{1}{m_{S_2}^2})   \bigg )  \bigg \rbrack^2 + 
L \leftrightarrow R  \Bigg \} ,   
%\label{eq:gemuKOVS}  
\label{eq:gemuK0VS}  
\end{eqnarray}
\begin{eqnarray}
&&\Gamma(B^0 \to e \tau ) = 
\frac{m_{B^0} f_{B^0}^2}{32\pi} \bigg (1-\frac{m_{\tau}^2}{m_{B^0}^2} \bigg )^2 
\Bigg \{ \bigg \lbrack  \frac{4 \pi\alpha_{st}}{m_V^2} ( - \, \lineup{m}_{B^0}+m_{\tau}/2) - 
\nonumber\\
&& - \frac{h_1 h_3}{2} \bigg ( m_{\tau} \langle \frac{1}{m^2_S} \rangle^{L} -
 \, \lineup{m}_{B^0} c s (\frac{1}{m_{S_1}^2} -  \frac{1}{m_{S_2}^2})   \bigg )  \bigg \rbrack^2 + 
L \leftrightarrow R  \Bigg \} ,     
%\label{eq:getauBOVS}  
\label{eq:getauB0VS}  
\end{eqnarray} 
where 
\begin{eqnarray}
&&h_p = -\sqrt{ 3/2} \frac{1}{\eta \sin\beta} (  m_{d_p} - m_{l_p} ) ,  
\label{eq:hconst}
\\
%\nonumber
&&\langle \frac{1}{m^2_S} \rangle^{L}= \frac{s^2}{m_{S_1}^2}+\frac{c^2}{m_{S_2}^2} \, , \,\,\,\, 
\langle \frac{1}{m^2_S} \rangle^{R}= \frac{c^2}{m_{S_1}^2}+\frac{s^2}{m_{S_2}^2}, 
\label{eq:1mS2LR}
\\
%\end{eqnarray}
%\begin{eqnarray}
&&\lineup{m}_{K^0}= m_{K^0}^2/(m_{s}+m_{d}) , \,\,\,\,\,\;\; \lineup{m}_{B^0}= m_{B^0}^2/(m_{b}+m_{d}) ,  
%\nonumber
\label{eq:mKmBT}
\end{eqnarray}
the form factor $f_{K^0}$ is defined as 
\begin{eqnarray}
&&\langle 0| \bar{s} \gamma^{\mu} \gamma^{5} d| K^0(p) \rangle = i \, f_{K^0} p^{\mu}, \;\;\;\;\; 
%\\   
%\label{eq:faK}
%\nonumber
\langle 0| \bar{s} \gamma^{5} d| K^0(p) \rangle = - \, i \,\, \lineup{m}_{K^0}  f_{K^0},  
\label{eq:fpK}
\nonumber
\end{eqnarray}
 $p_{\mu}$ is 4-momentum of the decaying meson 
(the same definition has been used for~$f_{B^0}$) 
and the relations 
$K^0=(\bar{s}d)$, \, $K^0_L=(\bar{s}d - \bar{d}s)/\sqrt{2}$, \, $ B^0=(\bar{b}d)$ 
have been taken into account.

In formulae~(\ref{eq:gemuK0VS}),~(\ref{eq:getauB0VS})  
we imply that     
\begin{eqnarray}
\Gamma(K^0_L \to e \mu ) &=& \Gamma(K^0_L \to e^- \mu^+ )+\Gamma(K^0_L \to e^+ \mu^- )= 
%\nonumber \\
% &=& 
2 \Gamma(K^0_L \to e^- \mu^+ ) ,
\nonumber
\\
%\label{eq:defgemuKO}  
%\end{eqnarray}
%\begin{eqnarray}
\Gamma(B^0 \to e \tau ) &=& \Gamma(B^0 \to e^- \tau^+ )+\Gamma(\tilde{B}^0 \to e^+ \tau^- )= 
2 \Gamma(B^0 \to e^- \tau^+ ) .  \hspace{6mm} 
%\label{eq:defgetauBO}  
\nonumber
\end{eqnarray}

It is interesting to note that the contributions of the scalar leptoquarks 
in~(\ref{eq:gemuK0VS}),~(\ref{eq:getauB0VS}) can have the opposite sign relatively to
vector leptoquark contribution, so that the destructive interference of these 
contributions can take place in the decays under consideration. The magnitude 
of the scalar leptoquark contributions depends on the scalar leptoquark masses 
on the masses of quarks and leptons and on the scalar leptoquark mixing angle 
in~(\ref{eq:mixS2}).

The interaction of the chiral gauge leptoquarks with fermions can be obtained 
from~(\ref{eq:lagrVdl}) by believing 
\begin{eqnarray}
&&(g^{L}_1)_{pi}= \frac{g^L_4}{\sqrt{2}}(K^{L}_2)_{pi}, \,\,\, (g^{R}_1)_{pi}= 0 ,   
%\nonumber \\
\label{eq:g1const}
\\
&&(g^{L}_2)_{pi}= 0 , \,\,\, (g^{R}_2)_{pi}= \frac{g^R_4}{\sqrt{2}}(K^{R}_2)_{pi}    
\label{eq:g2const}
\end{eqnarray} 
and takes the form  
\begin{eqnarray}
  \emph{L}_{Vdl} &=& \frac{g^L_4}{\sqrt{2}} (\bar{d}_{p \alpha} 
[ (K^L_2)_{pi}\gamma^{\mu}P_L] l_i)V^L_{\alpha \mu} +  
%\nonumber \\
% &+&  
\frac{g^R_4}{\sqrt{2}} (\bar{d}_{p \alpha} 
[ (K^R_2)_{pi}\gamma^{\mu}P_R] l_i)V^R_{\alpha \mu} + h.c. , \hspace{5mm} 
\label{eq:lagrVLRdl}
\end{eqnarray}
where  $g^{L}_4$, $g^{R}_4$ are the gauge coupling constants of the group~(\ref{eq:G4LR}) or~(\ref{eq:G4SLR}) 
which are related to strong coupling constant by the equation 
\begin{eqnarray}
g^L_4 g^R_4 / \sqrt{(g^L_4)^2+(g^R_4)^2} = g_{st} . 
\label{eq:gLgRgst}
\end{eqnarray}

As a result of calculations we have obtained the widths of 
the $K_L^0 \to e \mu , \; B^0 \to e \tau $ decays with account of the contributions 
of the chiral gauge leptoquarks $ V^L, V^R $ in the form  
\begin{eqnarray}
\Gamma(K^0_L \to e \mu ) =  \frac{m_{K^0}f_{K^0}^2m_{\mu}^2} {64\pi} 
\bigg (1-\frac{m_{\mu}^2}{m_{K^0}^2} \bigg )^2 
\bigg \lbrack \frac{(g^L_4)^4}{4m_{V^{L}}^4} |\kappa'^L_{21}|^2 + 
L \leftrightarrow R \bigg \rbrack , \hspace {8mm}   
\label{eq:gemuKOLR}  
\end{eqnarray}
\begin{eqnarray}
\Gamma(B^0 \to e \tau ) &=&  \frac{m_{B^0}f_{B^0}^2m_{\tau}^2} {32\pi} 
\bigg (1-\frac{m_{\tau}^2}{m_{B^0}^2} \bigg )^2 \times 
%\nonumber \\ 
%&& \times 
\bigg \lbrack \frac{(g^L_4)^4 }{4m_{V^{L}}^2} |k'^L_{31}|^2 + L \leftrightarrow R  \bigg \rbrack ,
\label{eq:etauBOLR}  
\end{eqnarray}
where the parameters 
\begin{eqnarray}
\kappa'^{L,R}_{ij} &=& (K^{L,R}_2)_{2i}\,\,\starupp{(K^{L,R}_2)_{1j}} - 
(K^{L,R}_2)_{1i}\,\,\starupp{(K^{L,R}_2)_{2j}} ,    
\label{eq:kLRijKO}  
\end{eqnarray} 
\begin{eqnarray}
%m^{L,R}_{ij} &=& \bar{m}_{B^0} k^{L,R}_{ij} + (m_{l^-_j}k'^{L,R}_{ij}+m_{l^+_i}k'^{L,R}_{ij})/2, 
%\label{eq:mLRij} \\ 
%k^{L,R}_{ij} &=& (K^{L,R}_2)_{3i}\,\,\starupp{(K^{R,L}_2)_{1j}}, 
%\label{eq:kLRijBO} \\  
k'^{L,R}_{ij} &=& (K^{L,R}_2)_{3i}\,\,\starupp{(K^{L,R}_2)_{1j}}  
\label{eq:k1LRijBO}  
\end{eqnarray}
account the effects of the fermion mixing in the leptoquark currents.

We have numericaly analysed 
the widths~(\ref{eq:gemuK0VS}),~(\ref{eq:getauB0VS}),~(\ref{eq:gemuKOLR}),~(\ref{eq:etauBOLR}) 
in dependence on the leptoquark masses with varying the values of the quark masses 
in the scalar leptoquark coupling constants and in~(\ref{eq:mKmBT}) ) . The gauge coupling constants we relate 
to $g_{st}(M_c)$ at $M_c=1000 \, TeV$, in the case of chiral gauge leptoquarks 
we assume also that $g^L_4=g^R_4 (=\sqrt{2}g_{st})$. 
For simplicity we assume $K^L_2=K^R_2=I$ and for definiteness we believe $\sin\beta = 0.7 . $   
We use the values  
of the form factors $f_{K^0}= 160 \, MeV, \,\, f_{B^0}=170 \, MeV$ and  
%the total widths  
%$\Gamma_{K^0_L}^{tot}$ and  $\Gamma_{B^0}^{tot}$ of $K^0_L$ and $B^0$ mesons 
%defined by 
the life times~\cite{PDG04} of $K^0_L$ and $B^0$ mesons.

\begin{figure}[htb]
%\vspace*{0.5cm}
\centerline{\epsfxsize=0.8\textwidth\epsffile{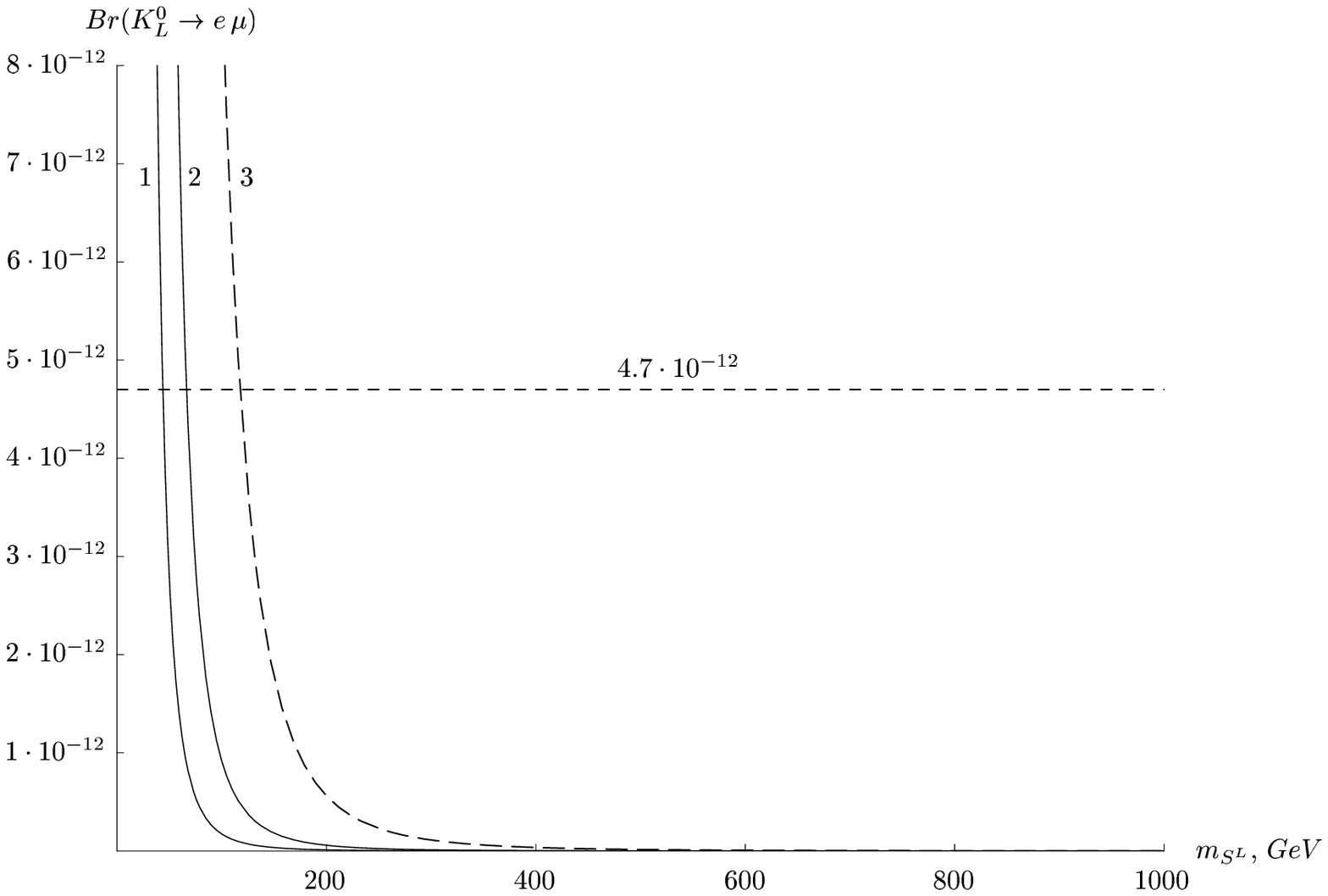}} 
%\vspace*{1mm}
\label{fig:BrKS} 
\caption{Branching ratio of $K^0_L \to e \mu$ decay 
in dependence on the  mass $m_{S^L}$ of the chiral scalar leptoquark $S^L$ for 
$1)m_d=11 \, MeV, \,\, m_s=175 \, MeV, \,\,\,\, $ 
$2)m_d=14 \, MeV, \,\, m_s=230 \, MeV, \,\,\,\, $ 
$3)m_d=22 \, MeV, \,\, m_s=350 \, MeV. $ }
\end{figure}

Fig.1 shows the branching ratio of $K^0_L \to e \mu$ decay 
in dependence on the  mass $m_{S^L}$ of the chiral scalar leptoquark $S^L$. 
The horizontal dashed line shows the experimental upper limit 
$Br(K^0_L \to e \mu)<4.7\cdot 10^{-12}$ \cite{PDG04}.  
The curves 1, 2 correspond to quark masses  
\begin{eqnarray}
 m_d=11 \, MeV, \,\, m_s=175 \, MeV ,  
\label{eq:mq1}                 
\\
 m_d=14 \, MeV, \,\, m_s=230 \, MeV    
\label{eq:mq075}  
\end{eqnarray}
which are taken at the  mass scale $ \mu=1 \, GeV$ 
and (approximately) at $\mu=750 \, MeV$ respectively. 
Strictly speeking one should to use the quark masses at the mass
of $K^0_L$-meson but this is a nonpertubative region and we can
not calculate them. Nevertheless one can hope that the quark
masses at $m_{K^0_L}$ are larger that those at 750 MeV and, hence
the contribution of the scalar leptoquark is also larger. To
illustrate how such contribution could affect the branching ratio of 
$K^0_L \to e \mu$ decay we
present also the curve 3 for the quark masses 
\begin{eqnarray}
 m_d=22 \, MeV, \,\,\, m_s=350 \, MeV 
\label{eq:mqx}  
\end{eqnarray}
which are chosen twice as large as those at $\mu=1 \, GeV$ . 
As seen in all three cases the lower mass limits for the chiral scalar leptoquark 
are small, below the mass limits from the direct search for scalar leptoquarks. 
The contributions of the leptoquarks of the scalar or pseudoscalar types are slightly 
larger but the corresponding mass limits are still below the direct 
scalar leptoquark mass limits. 

\begin{figure}[htb]
%\vspace*{0.5cm}
 \centerline{\epsfxsize=0.8\textwidth \epsffile{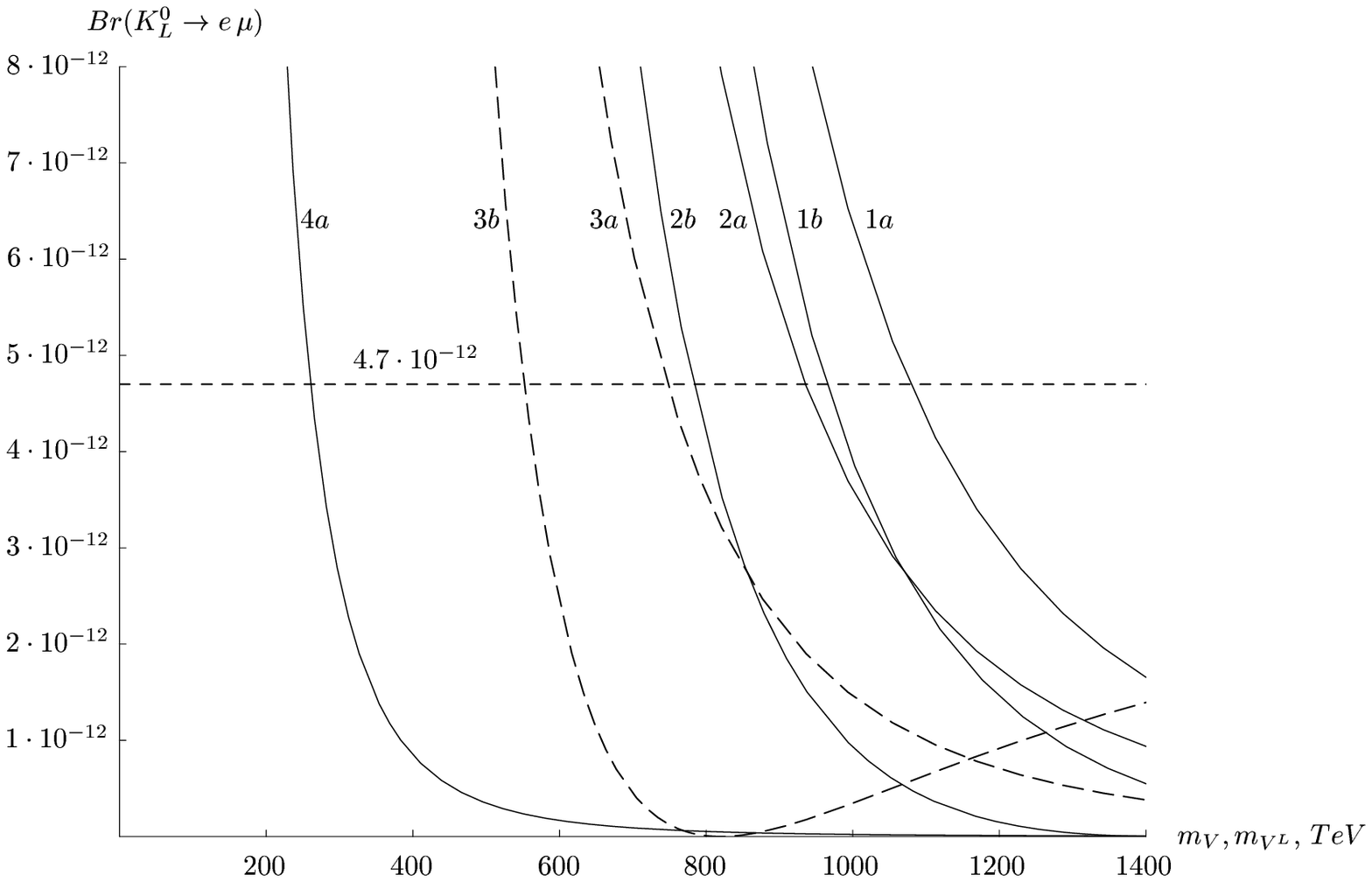}} 
%\vspace*{1mm}
\label{fig:BrKV} 
\caption{ Branching ratio of $K^0_L \to e \mu$ decay 
in dependence on the  mass $m_{V}$ of the vector leptoquark $V$ 
for the quark masses    
$m_d=11 \, MeV, \,\, m_s=175 \, MeV $ (curve $1$), 
$m_d=14 \, MeV, \,\, m_s=230 \, MeV $ (curve $2$),
$m_d=22 \, MeV, \,\, m_s=350 \, MeV $ (curve $3$)   
for the case $a)$ with neglect of the scalar leptoquark contribution     
and for the case $b)$ with account of the pseudoscalar leptoquark contribution                 
and on the chiral gauge leptoquark mass $m_{V^L}$ (curve $4a)$.} 
\end{figure}

The branching ratios of $K^0_L \to e \mu$ decay 
as the functions of the  masses $m_{V}, \, m_{V^L}$ of 
the vector and chiral gauge leptoquarks $V, \, V^L$ are shown in Fig.2. 
The curves 1, 2, 3 correspond to the quark 
masses~(\ref{eq:mq1}),~(\ref{eq:mq075}),~(\ref{eq:mqx}) respectively 
for the case $a)$ when only the vector leptoquark $V$ is taken into account 
with neglecting the scalar leptoquarks and for the case $b)$
with account of both the vector leptoquark $V$ and the pseudoscalar leptoquark $S^P$ 
with $m_{S^P}=250 \, GeV$.    
For the case $1 a)$ the lower vector leptoquark mass limit is of order
of $m_V> 1100 \, TeV $.

As seen, in these cases ( due to the destructive interference of the pseudoscalar 
and vector leptoquark contributions and due to the mass dependence in~(\ref{eq:mKmBT}) )   
the lower mass limit for the vector leptoquark can be decreased 
to about $m_V> 800 \, TeV $ or slightly below in dependence on the quark masses. 
 The curve $4 a)$ corresponds to case when only the chiral gauge leptoquark $V^L$ is taken into account. 
The lower mass limit for the chiral gauge leptoquark in this case  is of order
of $m_{V^L} > 260$ TeV. One can hope that the account of the scalar leptoquarks can result in 
further decreasing the mass limit for  the chiral gauge leptoquark.

As concerns the $B^0 \to e \tau$ decay the current experimental limit on its branching ratio 
$Br(B^0 \to e \tau)<5.3\cdot 10^{-4}$ \cite{PDG04} gives the relatively weak limits 
on the leptoquark masses. For example the lower  mass limits from $B^0 \to e \tau$ decay 
for the scalar leptoquarks are only of order of a few GeV, i.e. they lie essentialy 
below the mass limit from the direct search for scalar leptoquarks. 
The mass limits from $B^0 \to e \tau$ decay for the gauge leptoquarks are also weaker 
than those from $K^0_L \to e \mu$ decay nevertheless they are of interest 
as the new independent on $K^0_L \to e \mu$ decay mass limits.      

\begin{figure}[htb]
%\vspace*{0.5cm}
 \centerline{\epsfxsize=0.8\textwidth \epsffile{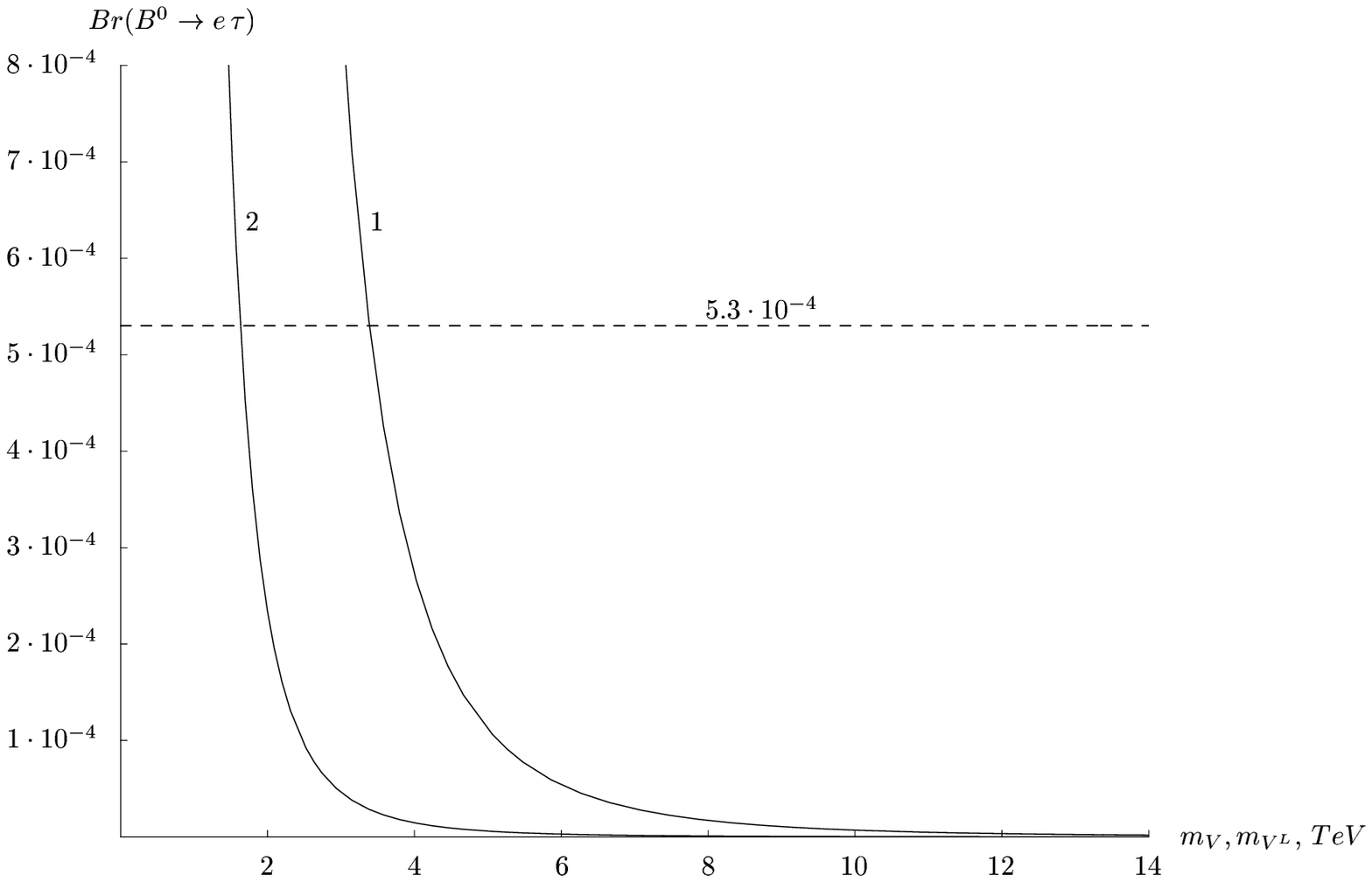}} 
%\vspace*{1mm}
\label{fig:BrBV} 
\caption{ Branching ratio of $B^0 \to e \tau$ decay 
in dependence on the  mass $m_{V}$ of the vector leptoquark $V$  (curve 1)   
and on the chiral gauge leptoquark mass $m_{V^L}$ (curve 2).} 
\end{figure}

The Fig.3 showes the branching ratios of $B^0 \to e \tau$ decay  
as the functions of the  masses $m_{V}, \, m_{V^L}$ of 
the vector and chiral gauge leptoquarks $V, \, V^L$.  
The curve 1 corresponds to the case when only the vector leptoquark $V$ 
is taken into account with neglecting the scalar leptoquarks.
In this case the lower mass limit for the vector leptoquark is of order
of $m_V> 3.4  \, TeV$.
 The curve 2 corresponds to the case when only the chiral gauge leptoquark $V^L$ 
is taken into account. 
The lower mass limit for the chiral gauge leptoquark in this case  is of order
of $m_{V^L} > 1.6 \, TeV$. Because the contributions into $B^0 \to e \tau$ decay from 
the scalar leptoquarks with masses allowed by their direct search are small 
the interference of the scalar and gauge leptoquark contributions 
in the $B^0 \to e \tau$ decay is negligible.

In conclusion one can say that the mass limits for the scalar leptoquark doublets 
from the current experimental bounds on the branching ratios 
of $K^0_L \to e \mu$ and $B^0 \to e \tau$ decays   
( in contrast to the corresponding mass limits for the gauge leptoquarks ) 
occur to be small, of order of or below the mass limits from the direct search 
for scalar leptoquarks. The search for these scalar leptoquarks at LHC 
as well the further search for the $ B^0$ leptonic decays of 
$ B^0 \to l^+_i l^-_j $ type are of interest.

The author is grateful to Organizing Committee of the
International Seminar "Quarks-2006" for possibility to participate
in this Seminar. The work was partially supported by the Russian
Foundation for Basic Research under the grant No 04-02-16517-a.

\end{document}